\documentclass[aps, pra, a4paper,showpacs,twocolumn, 10pt]{revtex4-1}
\usepackage{bbm, amsmath, amssymb, amsthm, bm,textcomp, nicefrac,geometry,dsfont,soul}
\usepackage{amsthm,amssymb,amsmath,graphicx,epsfig,color,verbatim,enumerate,amsthm}

\newcommand{\ket}[1]{\left|#1\right\rangle}

\newcommand{\bra}[1]{\left\langle #1\right|}

\newcommand{\proj}[1]{\ket{#1}\!\!\bra{#1}}
\newcommand{\mn}[1]{\langle #1 \rangle}

\newcommand\QFI{{\mathcal F}}

\newcommand{\bea}{\begin{eqnarray}}
\newcommand{\eea}{\end{eqnarray}}

\newcommand\ii{\mathrm{i}}
\def\c{\textrm{c}}
\def\s{\textrm{s}}

\def\tr{\mathrm{tr}}

\makeatletter
\newtheorem*{rep@theorem}{\rep@title}
\newcommand{\newreptheorem}[2]{%
\newenvironment{rep#1}[1]{%
 \def\rep@title{#2 \ref{##1}}%
 \begin{rep@theorem}}%
 {\end{rep@theorem}}}
\makeatother

\newreptheorem{theorem}{Theorem}

\newcommand{\eins}{\mathbbm{1}}
\newcommand{\be}{\begin{equation}}
\newcommand{\ee}{\end{equation}}

\begin{document}
\title{Improved sensing with a single qubit}
\author{P. Sekatski$^1$, M. Skotiniotis$^2$ and W. D\"ur$^1$}

\affiliation{$^1$ Institut f\"ur Theoretische Physik, Universit\"at
  Innsbruck, Technikerstr. 21a, A-6020 Innsbruck,  Austria}
\affiliation{$^2$ F\'{\i}sica Te\`{o}rica: Informaci\'{o} i Fen\`{o}mens Qu\`{a}ntics, Departament de F\'{\i}sica, 
Universitat Aut\`{o}noma de Barcelona, 08193 Bellaterra, Spain }
\date{\today}

\begin{abstract}
We consider quantum metrology with arbitrary prior knowledge of the parameter. We demonstrate that a single sensing two-level 
system can act as a virtual multi-level system that offers increased sensitivity in a Bayesian, single-shot, metrology scenario and 
that allows one to estimate (arbitrary) large parameter values by avoiding phase wraps. This is achieved by making use of 
additional degrees of freedom or auxiliary systems not participating in the sensing process. The joint system is manipulated by 
intermediate control operations in such a way that an effective Hamiltonian, with an arbitrary spectrum, is generated that 
mimics the spectrum of a multi-level system interacting with the field. We show how to use additional internal degrees of freedom 
of a single trapped ion to achieve a high-sensitivity magnetic field sensor for fields with arbitrary prior.
\end{abstract}
\maketitle

\paragraph*{Introduction.---}
Quantum metrology deals with one of the most fundamental problems in science---the measurement or estimation of an 
unknown physical quantity.  In the widely studied case of phase estimation it was found that quantum mechanics allows for a 
significant improvement in the achievable precision over classical schemes. This applies both to the local estimation 
scenario~\cite{GLM04, Giovannetti2006, Giovannetti2011}, where the experiment is repeated sufficiently many times without any 
modification or the phase is already known to lie within a localized set of values, and to reference frame alignment 
scenario~\cite{Berry00,Chiribella04}, where the phase is totally unknown prior to the estimation and a single measurement is 
performed. In this letter we consider the problem of frequency estimation, a variant of the phase estimation scenario, which, 
however, reduces to neither of the two scenarios above. This is due to the additional freedom of choosing the interaction time at 
each run, and to the occurrence of phase ambiguities (known as phase-wraps) due to the periodicity of unitary evolution.

Here we propose a simple scheme that uses a single sensing particle and additional auxiliary systems 
(or internal degrees of freedom) that do not participate in the sensing process together with (fast) quantum control. 
We show that one can mimic the evolution of an effective $n$-level system with an arbitrary spectrum of constant spectral 
radius and demonstrate that such an effective $n$-level system offers significant advantages in different metrological tasks.
After providing an exact solution for the optimal states and measurements that achieve the 
minimum average squared error for a two-level probe (a qubit), we 
show how to make use of additional degrees of freedom of a single system---that are not affected by the evolution---to engineer 
an effective $n$-level system.  We then show how such spectral engineering can be done on-the-fly, allowing for a prolonged 
interrogation time of the protocol and avoiding multiple measurements and re-preparation of the states.  
In the case where multiple sensing systems subjected to local Hamiltonian evolution our spectral engineering technique can 
also be used to improve sensing as we explicitly show for the case of two qubits.  
Lastly, we outline precisely how our techniques can be applied for the case of Bayesian magnetometry using a 
single trapped-ion. 


We remark that in the absence of noise quantum control does not allow to increase the quantum 
Fisher information (QFI), since it only depends on the extremal eigenvalues of the Hamiltonian.  However, as we demonstrate, 
the QFI is never the directly relevant quantity in the noiseless Bayesian frequency estimation as 
it corresponds to the second order expansion of the precision in the inquiry time, and it is always better to increase the latter beyond this regime. 

\paragraph*{Background and setting of the problem.---}
\label{sec.Background}
Consider a single sensing system whose evolution is governed by $H_S=\omega\, h$, where the frequency
$\omega$ is known with  prior probability $p_0(\omega)$  with variance $V_0$.  The goal is to estimate 
the value of $\omega$ by preparing the system in an initial state, $\rho =\proj{\psi}$, allowing it to freely evolve under the 
unitary $U=\exp{(-i \,t \,H_S)}$ for a time $t$ before measuring the state of the system, gaining some information about 
$\omega$, and updating the corresponding probability density.  We note that one  can also assume without loss of generality 
that the prior probability density has zero mean. The precision of the estimation is quantified by the mean posterior variance,  or 
mean squared error (MSE) $\mn{V(\omega)}_t\equiv \sum_m p_m \textrm{Var}\left(p(\omega|m)\right)$,
where $p(\omega|m)$ is the posterior probability density 
upon obtaining outcome $m$ (and we also define the estimators $\omega_m = \int \omega \, p(\omega|m)d\omega$).

Consider the decrease of the MSE with time $t$. For well-behaved priors it can be lower-bounded using the Bayesian 
Cram\'{e}r-Rao inequality~\cite{Gill:95}
\be
\mn{V(\omega)}_t\geq \frac{1}{\mathcal{I}(p_0(\theta))+ t^2 \QFI_h(\rho)},
\label{eq.BCRB}
\ee 
where $\QFI_h(\rho)$ is the QFI of the state $\rho$ with respect to the generator $h$, and 
$\mathcal{I}(p_0(\theta))\equiv \int \frac{(\partial_\omega p_0(\omega))^2}{p_0(\omega)} d\omega$ is the (time independent) 
Fisher information of the prior.  Moreover, for short times the average variance decreases quadratically with $t$ 
with a pre-factor given by
\be
\mn{V(\omega)}_t = V_0- t^2\,  V_0^2 \mathcal{F}_h(\rho) + \mathcal{O}(t^3),
\label{eq.Vardecrease}
\ee
as we show in App.~\ref{app:QFI}.

However, $\mn{V(\omega)}_t$ can not decrease indefinitely with $t$.  Indeed, Holevo's theorem~\cite{Holevo:73} places an 
upper bound on the mutual information $I(m\!:\!\omega)$ between the measurement 
outcome and the parameter, based on the maximal amount of information that the output state can encode.  As we show in App.~\ref{app:Entropic} Holevo's bound implies the 
following bound on the average variance
\be
\mn{V(\omega)}_t \geq \frac{1}{d^2}\frac{e^{2 H(p_0(\omega))}}{2\pi e},
\label{eq.Holevo}
\ee
where $H(p_0(\omega))$ is the Shannon entropy of the prior.  For a Gaussian prior of width $\sigma$ 
Eq.~\eqref{eq.Holevo} simplifies to $\mn{V(\omega)}_t \geq\left(\frac{\sigma}{d}\right)^2$.  Hence, the 
number of non-degenerate levels of the probe state (with respect to the eigenbasis of $H_S$) directly bounds the attainable 
precision, regardless of the inquiry time $t$. To determine the optimal time to perform the measurement we must first determine the optimal measurement.

For von Neumann measurements, i.e., POVMs comprised of orthogonal rank-one projectors $E_m$, it was shown 
in~\cite{Personick:71} that the measurement minimizing the MSE has to satisfy 
\bea
\Gamma_t S_t + S_t\, \Gamma_t =2 \eta_t, 
\label{eq:S}
\eea 
where $\Gamma_t =\int e^{- i t H_S}\, \rho\, e^{i t H_S}  p_0(\omega)\, \mathrm{d}\omega$,
$\eta_t =\int  e^{- i t H_S}\,\rho\, e^{i t H_S}\,  \omega\, p_0(\omega)\, \mathrm{d}\omega$ and $S_t = \sum_m \omega_m E_m$.  
In addition, the average variance attained by such a measurement is given by 
\bea
\langle V( \omega) \rangle_t =V_0 - \tr\, \eta_t S_t.
\label{eq.SLD}
\eea
Moreover, it was later proven that von Neumann measurements are generally optimal for the MSE figure of 
merit~\cite{Macieszczak:14}, so that the solution $S_t=\sum_m \omega_m E_m$ of \eqref{eq:S} contains the optimal 
measurement and estimators, and provides the best attainable MSE via Eq.~\eqref{eq.SLD}.

Notice for any $H_S$ with a finite gap the decrease in the average variance vanishes in the limit $t\to \infty$ whenever the prior has a bounded spectrum.
This follows from the fact that $\lim_{t\to \infty} \eta_t \to 0$. Intuitively this happens because beyond a certain time phase wrap ambiguities render the 
information obtained via the measurement useless for the MSE. Hence, there exists an optimal time, 
$t_\textrm{max} =\textrm{arg}\min_t (\mn{V(\omega)}_t)$ for which the error is minimized, and after which 
the phase begins to wrap due to the periodicity of the unitary evolution $e^{i t H_S}$.  

Now consider the case where the probe is a single qubit and $h=\frac{1}{2} \sigma_z$. The optimal initial state of the probe 
always lies on the equator of the Bloch sphere (see App.~\ref{app:equator}) and, due to the symmetry of the problem with 
respect to simultaneous rotations about the $z$-axis, one can choose it to be 
$\ket{+} =\frac{1}{\sqrt{2}}\left(\ket{0}+\ket{1} \right)$. This results in 
\bea
\Gamma_t =
\frac{1}{2}{\small 
\left(\begin{array}{cc}
1&\tilde p_0(t) \\ 
\tilde p^*_0(t) & 1 
\end{array}\right)},\; 
\eta_t =
\frac{1}{2}{\small 
\left(\begin{array}{cc}
0&\ii\, \tilde p_0'(t) \\ 
-i\tilde p_0^{'*}(t) & 0 
\end{array}\right)},
\eea
where $\tilde p_0(t) = \int e^{i t \omega} p_0(\omega) d\omega$, and $\tilde p_0'(t) = \frac{\mathrm{d}}{\mathrm{d}t} \tilde p_0(t)$. 
For this simple case an exact analytical solution for both the optimal measurement strategy, $S_t$, as well as the optimal time 
can be found in App.~\ref{app:general prior}. For a symmetric prior $p_0(\omega)=p_0(-\omega)$, the solution of Eq.~\eqref{eq:S} is particularly simple;
$S_t = 2 \, \eta_t$, and the average variance reads
\be
\Delta \mn{V(\omega)}_t \equiv \langle V(\omega) \rangle_t - V_0(\omega) = \tr\, \eta_t S_t =  \left(\tilde p'(t)\right)^2,
\label{eq:ss variance}
\ee
making it easy to determine $t_\textrm{max}$.


A good strategy, then, is to prepare the system in the state $\ket{+}$, subject it to the noiseless unitary evolution and measure
the subsequent state shortly before $t_\textrm{max}$.  Depending on the measurement outcome, $m$, we
update our knowledge of the parameter to $p_1(\omega)=p(\omega|m)$, and repeat the 
protocol again for the updated distribution~\footnote{Note that also weak measurements, without a fresh preparation procedure 
are in principle possible.}. Though this strategy is easy to formulate, finding the optimal performance and 
sequence of optimal measurement settings is infeasible beyond a certain number of steps, as one has to keep track of the exact 
form of the prior at each step and for each measurement outcome (just keeping track of the variance is insufficient). This is why 
to assess the performance one is constrained to numerical search~\cite{Sergeevich:11}. 
In App.~\ref{app:sequential} we find the optimal performance under the wrong assumption that the prior knowledge stays 
Gaussian at every step. Our result suggests that after a large number of steps such a recursive strategy achieves the optimal 
scaling precision $\langle V(\omega)\rangle_t \propto 1/T^2$, where $T$ is the total running  time. 
   
However, the practical applicability of such sequential schemes may be strongly limited, as they require multiple measurements 
and preparation procedures. In most set-ups state preparation and measurements are rather time 
consuming, and hence cannot be neglected. For instance, for ion traps cooling of ions to the motional ground state as well as measurements 
in an ion trap set-up take place on a much longer time scale than unitary control pulses. In practice, this limits the minimal 
evolution time, and the number of intermediate measurements.  We now present an alternative scheme that requires only one 
state preparation step at the beginning and one measurement step at the end, whereas at intermediate times only some limited 
unitary control is required.

\paragraph*{Degeneracy lifting.---}
\label{sec.degeneracy}
We consider a single sensing qubit whose evolution is governed by $H_S$.  Moreover, we assume that there are additional 
auxiliary systems or internal degrees of freedom available that are not affected by $H_S$ and that we have fast control over the 
sensing plus ancilla systems or additional degrees of freedom. We describe the latter as a $n$-level system, so that the total 
Hamiltonian is given by $H_{SA}=H_S \otimes \eins_A$, and a basis of the joint system is given by 
$\{|\psi_{j,k}\rangle=|j\rangle \otimes |k\rangle\}$ with $j\in \{0,1\}, k\in \{1,2,\ldots n\}$. The total Hamiltonian still has only two 
distinct eigenvalues, however the eigenvectors are $n$-fold degenerate, $\{|\psi_{0,k}\rangle\}$ and $\{|\psi_{1,k}\rangle\}$.

We first show that for such a set-up, with one sensing qubit plus an $n$-level auxiliary system, and any fixed evolution time 
$t$ one can obtain an effective $2n$-level system that evolves under an effective Hamiltonian with an arbitrary spectrum 
$\{\lambda_\ell\}$ with $2n$ eigenvalues, where $|\lambda_\ell| \leq 1$, 
$H_{\rm eff} = \frac{1}{2} \omega \sum_{\ell=1}^{2n} \lambda_\ell |\ell\rangle\langle \ell|$. That is, an $n$-level state
\be
\label{psin}
\ket{\psi} = \sum_{\ell=1}^{2n} c_\ell \ket{\ell}
\ee
evolves to
$|\psi_t\rangle = \sum_{\ell=1}^{2n} c_\ell e^{-i\frac{1}{2}\omega \lambda_\ell t} \ket{\ell}$. In addition, any state 
$\ket{\psi}$ with real positive coefficients $c_\ell$ can be mimicked within the same procedure~\footnote{the restriction to real 
positive coefficients is not necessary, but it does not affect the metrological performance} as we now explain. 

This is achieved by preparing the probe in initial state $\c_0\ket{0,1}+ \s_0 \ket{1,1}$. At times $t_{j,k}\leq t_{j,k+1}$
control pulses $U_{j,k}$ rotating between the levels $\ket{j,1}$ and $\ket{j\oplus1,k}$ are applied. In the basis
$\{\ket{j,1},\ket{j\oplus1,k} \}$ the control unitary reads
$
U_{j,k}= {\small
\left(\begin{array}{cc}
\cos(\vartheta_{j,k})& -\sin(\vartheta_{j,k})\\
\sin(\vartheta_{j,k}) & \cos(\vartheta_{j,k})
\end{array}\right)}.
$
At time $t$ each level $\ket{j,k}$ picks the phase $\exp\left(- t \frac{\ii \omega}{2} \lambda_{j,k} (1-2j) \right)$ with
$\lambda_{j,k}=1-2\frac{t_{j,k}}{t}$, resulting from the sum of phases picked before and after the time $t_{j,k}$ when the control
was applied (to shorten the notation we formally introduced $t_{j,0}=0$). Consequently the final state reads
\be
\ket{\psi}_t = \sum_{j,k} c_{j,k} e^{-\ii \frac{\omega}{2} \lambda_{j,k} t}\ket{j,k},
\label{eq.state}
\ee
with $c_{0,k} = \s_0 \sin(\vartheta_{0,k}) \prod_{j<k}\cos (\vartheta_{0,j}) $ and $c_{1,k} = \c_0 \sin (\vartheta_{1,k}) \prod_{j<k} \cos(\vartheta_{1,j}) $ for all
$k>1$, and $c_{0,1} = \c_0 \prod_{j} \cos(\vartheta_{1,j})$ and $c_{1,1} = \s_0 \prod_{j} \cos_(\vartheta_{0,k})$. It is easy to see that
with this parametrization the coefficients $\{ c_{j,k}\}$ span all normalized vectors with positive coefficients. Remark that any
adaptive sequential scheme with $N$ measurement steps and predefined running times for each step corresponds to a particular case of the general strategy above, with a $H_\text{eff}$ that has $2^N\!$ levels (with additional constraints on the spectrum), a flat input state $\ket{\psi}=\frac{1}{2^{N/2}}\sum_\ell \ket{\ell}$ and the final measurement that is also constrained. 

\begin{figure}[!ht]
\centering
\includegraphics[width=0.98\columnwidth]{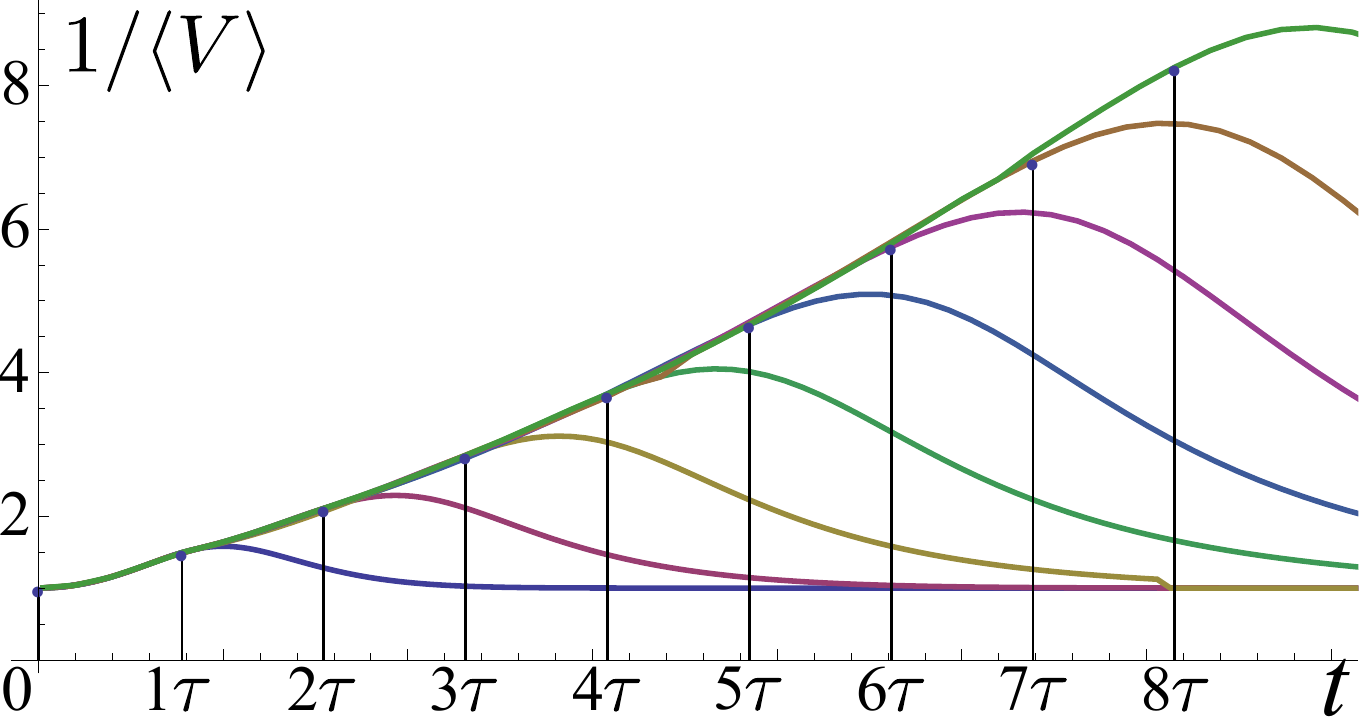}
\caption{Optimal variance attained by an equally gapped $n$-level state for  $2\leq n \leq 9$. The dots correspond to the times 
where the controls are turned on to either measure out the state or prepare the optimal state for the next time.}
\label{fig:OptimalVariance}
\end{figure}

We numerically evaluate the performance of the above strategy for the case of a symmetric prior density and a state $\ket{\psi}$ 
with real coefficients. For this case one finds that $\Gamma_t$ is a real symmetric matrix, whereas $\eta_t= \ii \eta_A$ with 
$\eta_A$ a real antisymmetric matrix. Solving Eq.~\eqref{eq.SLD} to find the optimal measurement yields $S=\ii A$ with real 
antisymmetric $A$.  To find the optimal pair $\{\ket{\psi}, A\}$ we numerically minimize the variance decrease for a fixed time
$t$ and number of non-degenerate levels $n$.  Fig.~\ref{fig:OptimalVariance} shows the performance of our protocol for up to 
$n=9$ equally spaced levels. To simplify the analysis from now on we shall only consider the case of engineering Hamiltonians 
with an equally gapped spectrum. This simplification is based on numerical evidence suggesting that equally gapped spectra are 
nearly optimal in time regimes where the average variance grows, i.e., before one encounters the phase wraps.  However, we 
stress that the degeneracy lifting protocol described above allows one to engineer any spectrum.  

We would like to note that fast control techniques to boost estimation precision using a single spin only were also considered 
in~\cite{Goldstein:11}. However, contrary to our approach,  the surrounding auxiliary particles were also assumed 
to be subjected to the evolution.

\paragraph*{On-the-fly spectral engineering.---}
\label{sec.engineering}

As mentioned earlier in order to achieve the the maximum gain in the average variance one must measure the system 
at the right time $t_{\mathrm{max}}$, before one encounters the phase wrap. Reciprocally for any fixed measurement time 
there exists an optimal spectrum and state minimizing the MSE that can be attained with the strategy above. However this 
scheme requires the measurement time  to be known before the sensing process even starts. We now show how the same can be
achieved with an \emph{on-the-flight} scheme, where in order to avoid phase wraps and repel the bound of Eq. \eqref{eq.Holevo} additional virtual levels are introduced during the sensing process, keeping the extractable MSE nearly optimal for all times.


As suggested by Eq.~\eqref{eq.Vardecrease} and can be seen explicitly form Fig.~\ref{fig:OptimalVariance}, no additional levels are needed in order to minimize the average variance for $t\leq \tau$. In the next time-slot, between $\tau$ and $2\tau$ ,the three-level strategy performs optimally.  But from $2\tau$ to $3\tau$ the performance of the three-level state starts to decay, however it is enough to go for the four-level one in order to recover the optimal MSE. This is due to the fact that as time increases, one needs to populate more and more intermediate levels in order to avoid phase wraps.  Indeed, the results of Fig.~\ref{fig:OptimalVariance} show that the number of intermediate levels required to maintain an optimally decreasing average variance grows linearly with time (such that the spectral gap of the effective Hamiltonian times the inquiry time $t$ stays constant). 
%

Consider a stroboscopic scenario where the controls are performed at times which are multiples of $\tau$ (for a Gaussian prior 
we heuristically determined a good time step duration to be around $\tau=0.775/\sigma$). One initially prepares the state 
$\ket{+,0}$, i.e., an equal superposition of the $\pm1/2$ eigenstates of the sensing Hamiltonian and ground state in the auxiliary 
degree of freedom. The state evolves for a time $\tau$ and the two levels pick a phase $e^{\pm \ii \tau \omega/2}$. At this point 
in time the state is optimal for sensing---it attains the minimal variance among all states given the evolution time $\tau$---letting 
this state evolve for a longer time results in a reduction in precision (see Fig.~\ref{fig:OptimalVariance}). Thus,  at time $\tau$ 
one intrudes an additional auxiliary level, by implementing the degeneracy lifting procedure described above and 
maps the state $\ket{\Psi_1}=\sqrt{0.5}e^{+ \ii \tau \omega/2}\ket{\uparrow,0}+\sqrt{0.5}e^{- \ii \tau \omega/2}\ket{\downarrow,0}$ 
to
\bea
\ket{\Phi_1}=e^{+ \ii \tau \omega/2}(\sqrt{0.3}\ket{\uparrow,0}+\sqrt{0.2}\ket{\downarrow,1})\nonumber\\ +e^{- \ii \tau \omega/2}
(\sqrt{0.3}\ket{\downarrow,0}+\sqrt{0.2} \ket{\uparrow,1}).
\label{eq.thirdlevel}
\eea

\begin{figure}[!ht]
\centering
\includegraphics[width=0.98\columnwidth]{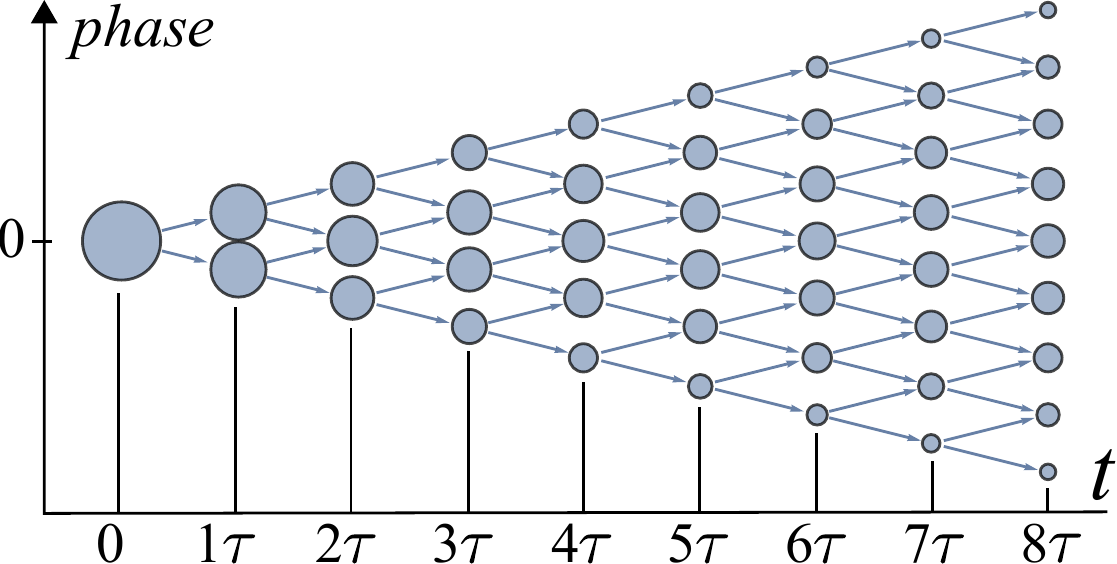}
\caption{Optimal states for each time step. The ordinate depicts the total phase acquired by each level at a given time, and the area of each circle corresponds to the weight of each level for the optimal states.}
\label{fig:Opt}
\end{figure}

At time $2\tau$ the state evolves to $\ket{\Psi_2}=e^{+ 2 \ii \tau \omega/2} \sqrt{0.3}\ket{\uparrow,0}+\sqrt{0.4} \ket{+,1} +e^{- 2 \ii 
\tau \omega/2} \sqrt{0.3} \ket{\downarrow,0}$, which is the optimal three-level state for this time. This procedure can be carried 
on by introducing an additional auxiliary level at each time step and mapping $\ket{\Psi_k}$ onto $\ket{\Phi_k}$ chosen such that 
it evolves to the optimal state $\ket{\Phi_{k+1}}$ for the next time step. In order to be able to do this it is necessary that the 
weights of all levels of the state $\ket{\Psi_{k+1}}$ correspond to a reshuffling between adjacent levels of the state  
$\ket{\Psi_{k}}$. Fig.~\ref{fig:Opt} depicts the phase evolution of different levels and shows how the above procedure works for constructing the optimal states at each time step for up to nine additional intermediate levels. The attained precision is given in Fig.~\ref{fig:OptimalVariance}.

Whilst we did not perform the numerical optimization of the states for more intermediate levels, we note that this can be easily 
done using more powerful numerical techniques such as those in~\cite{Rafal:11, Macieszczak:14}.  Alternatively, we note that
the sine states~\cite{Summy:90, Berry00} known to attain the optimal scaling of precision with time~\cite{Knysh:14,Sekatski:16} satisfy the weight reshuffling property and are thus achievable via this strategy.

\paragraph*{Multi-qubit systems.---}
\label{sec.multiqubits}

We now turn to multi-qubit systems and show that intermediate local 
control also allows one to increase the effective number of levels, even 
without using additional degrees of freedom or auxiliary systems. The spectrum of an 
$N$-qubit system with respect to the Hamiltonian $H=\omega\sum_{j=1}^N \sigma_z^{(j)}$ has 
$N+1$ distinct eigenvalues $\lambda_k =k$. The corresponding eigenstates 
are given by permutations of the state $|1\rangle^{\otimes 
k}|0\rangle^{\otimes N-k}$. For each $\lambda_k$, there are $\binom{N}{k}$ 
eigenstates $|\chi_{k,l}\rangle$. This degeneracy can be lifted by picking two states, 
$|\chi_{k_1,l_1}\rangle$ and $|\chi_{k_2,l_2}\rangle$ and swapping them 
at time $x T$ of the total evolution time $T$. This results in two 
new effective eigenvalues $x \lambda_{k_1} + (1-x) 
\lambda_{k_2}$ and $(1-x) \lambda_{k_1} + x \lambda_{k_2}$. 
In principle, any eigenvalue can be reached in this way. However, in this case there are limitations
on the effective spectrum one can achieve, as the number of degenerate eigenstates is different for different eigenvalues. 
For $N$ qubits one can achieve an arbitrary spectrum of exponentially many levels, whose spectral radius, however, is reduced 
by a constant. This is done by 
ignoring the outer parts of the spectrum, where the number of degenerate 
eigenstates is small, and only using the middle part of the spectrum 
where many degenerate eigenstates are available. When using only 
$\{\lambda_k\}$ for $N/4 \leq k \leq 3N/4$, the spectral width is 
reduced by a factor of $1/2$, and the number of available levels is 
given $\binom{N}{N/4} \approx exp(0.56 N)$.

By way of example, consider the simplest non-trivial case of two spin-$1/2$ particles where the overall spectrum of the 
Hamiltonian is $\{1,0,-1\}$ with a doubly degenerate $0$ eigenvalue. The use of dynamical decoupling allows to lift the 
degeneracy of the $0$-level at the price of reducing the spectral range, this allows to decrease the minimal attainable MSE by a 
factor of $1.36$ (see App.~\ref{app:2spins}).

\paragraph*{Noisy estimation.---}
Thus far our treatment assumes a noiseless Bayesian estimation scenario.  However, the 
techniques outlined so far can be used in conjunction with fast control and error-correction~\cite{Sekatski:16}, which serve to 
combat the effects of noise.  For example, by engineering a spectrum such that each distinct level is doubly 
degenerate~\footnote{This can be easily achieved, for example, by coupling all auxiliary degrees of freedom such that the 
effective Hamiltonian is non-degenerate and then introducing a single ancilla qubit that does not participate in the sensing 
process} one can implement the error-correcting schemes of~\cite{Sekatski:16} to eliminate all rank-one Pauli noise processes 
that act on the single sensing system. For general noise acting on the sensing system, part of the noise can still be eliminated 
with fast control and/or error-correction.  Uncorrectable noise will effectively act as correlated noise acting on the effective 
$n$-level system

\paragraph*{Realization with trapped ions}
\label{sec.ions}

A possible experimental implementation of our techniques can be realized in ion-trap set-ups to sense magnetic fields. There the 
internal electronic degrees of freedom of a single ion, $|g\rangle, |e\rangle$, form the bare qubit, and the motional degrees of 
freedom of the ion in the trap, labeled by $|k\rangle$, $1 \leq k \leq n$, provide the auxiliary degrees of freedom. Alternatively 
one can also use more of the electronic levels of the ion that are in priniciple accessible.  The 
Hamiltonian describing the evolution of both electronic and motional degrees of freedom is given by 
$H=\frac{B}{2}\sigma_z\otimes \eins$, where $\sigma_z=  |g\rangle\langle g| - |e\rangle\langle e|$ ($\eins$) acts on the 
electronic (motional) degrees of freedom respectively.  Thus, the total Hamiltonian contains just two distinct eigenvalues and, in 
principle, an infinite number of degenerate states for each eigenvalue, which correspond to the motional degrees of freedom of 
the ion in the trap.  

In order to engineer an effective Hamiltonian with an arbitrary spectrum one needs to be able to couple the electronic and 
motional degrees of freedom of the ion.  Moreover, in order to implement the protocols described above, 
one needs to be able to prepare arbitrary states of both the electronic and motional degrees of freedom as well as be able to 
perform arbitrary measurements on such states.  In App.~\ref{app:ions} we provide a detailed description of the coupling 
pulses, as well as the operations required to prepare such arbitrary states and measurements.  We stress that all operations 
required belong to the standard repertoire of ion trap quantum information processing~\cite{Alonso:13, Alonso:16, Ortiz:16}.
\paragraph*{Conclusion}
\label{sec.conclusion}

We have demonstrated a singe sensing system (a qubit), along with additional degrees of freedom, can be transformed into an 
effective multi-level system that offers increased sensitivity in a single parameter, noiseless Bayesian estimation scenario.  The 
engineering of such an effective multi-level system can be accomplished on the fly, by suitably coupling the qubit and auxiliary 
degrees of freedom at the appropriate times.  Furthermore, we have demonstrated how such on-the-fly engineering can be 
accomplished in ion trap experimental set-ups in order to perform high 
precision sensing of magnetic fields. 
 
We believe that our techniques can be readily applied to multi-parameter estimation problems, such as the precise estimation of 
an orthogonal triplet of spatial directions.  More importantly,
we believe that our techniques are highly relevant in the construction of atomic clocks, where the ambiguity of phase wraps is a 
hinderance to the stability of atomic clocks.

\textit{Acknowledgements.---}This work was supported by the Austrian Science Fund (FWF): P24273-N16, P28000-N27, SFB 
F40-FoQus F4012-N16, the Swiss National Science Foundation grant P300P2\_167749, Spanish MINECO  FIS2013-40627-P, 
and Generalitat de Catalunya CIRIT  2014 SGR 966.GR 966.

\appendix

\subsection{Recovery of the QFI for short times}

\label{app:QFI}
Consider the Taylor expansion of Eq.~(4) in terms of $t$.  The two operators are given by
\bea
\Gamma_t =& \sum_{n=0}^\infty \frac{(-\ii t)^n}{n!} \mn{\omega^n} [h,\rho]^{(n)}=\rho + \mathcal{O}(t^2)\\
\eta_t =& \sum_{n=0}^\infty \frac{(-\ii t)^n}{n!} \mn{\omega^{n+1}} [h,\rho]^{(n)}=- \ii t V_0(\omega) [h,\rho]+ \mathcal{O}(t^2),\nonumber\\
\eea 
where $[h,\rho]^{(n)}=[h,...,[h,\rho]...]$ is the nested commutator with $h$ appearing $n$ times.
This implies for the optimal measurement strategy
\be
S_t = t V_0(\omega) \mathcal{L}_{h}(\rho) + \mathcal{O}(t^2),
\ee
where $\mathcal{L}_{h}(\rho)$ is the symmetric logarithmic derivative~\cite{BC94} given by
\be
\rho \mathcal{L}_{h}(\rho)  + \mathcal{L}_{h}(\rho)  \rho = -2 \ii [h,\rho],
\ee
yielding
\be
\Delta V_t = \tr \,S_t \eta_t = t^2 V_0(\omega)^2 \mathcal{F}_{h}(\rho) + \mathcal{O}(t^3)
\ee
where $\mathcal{F}_h(\rho)$ is the quantum Fisher information.

\subsection{Entropic bound on the MSE}
\label{app:Entropic}
We now use Holevo's theorem~\cite{Holevo:73} to derive a lower bound on the mean average variance.  Holevo's theorem
establishes a relation between the mutual information of the 
measurement outcomes and the parameter encoded in a quantum state $I(m:\omega)$, and the Von Neumann entropy of the 
mean state $S(\Gamma_t)$
\be
I(m:\omega) \equiv H\left(p_0(\omega)\right)- \mn{H(m:\omega)} \leq S(\Gamma_t),
\label{Holevo}
\ee
where $H\left(p_0(\omega)\right)=H_0$ is the Shanon entropy of the prior distribution and 
$\mn{H(m:\omega)}=\sum_m p_m H\left(p(\omega|m)\right)$ is the mean Shanon entropy of the posterior. In addition, a trivial 
upper bound for the right hand side of Eq.~\eqref{Holevo} is given by the dimension, $d$, of the system, 
$S(\Gamma_t)\leq \ln(d)$. Rewriting Eq.~\eqref{Holevo} 
\be
\mn{H(m:\omega)} \geq H_0 -\ln(d),
\ee
allows us to obtain a bound for the MSE of the posterior distribution.

First notice that among all probability distributions with the same entropy the Gaussian has the least variance. Hence, for any 
distribution $p(\omega)$ the following inequality holds
\be
 \frac{1}{2} \ln\Big(2\pi e \textrm{Var}\big(p(\omega)\big)\Big)  \geq H\big(p(\omega)\big)  
\label{bound}
\ee
since $\ln\Big(\sqrt{2\pi e} \, \sigma \Big)$ is the entropy of a Gaussian with variance $\sigma^2$. As the logarithm is a concave 
function, the same bound in Eq.~\eqref{bound} holds for the mean variance $\mn{V(\omega)}$ and mean entropy 
$\mn{H(m:\omega)}$ over an ensemble of posterior distributions $\{\Big(p_m,p(\omega|m)\Big)\}$. Consequently, one obtains
\be
\mn{V(\omega)}\geq \frac{e^{2\mn{H(m:\omega)}}}{2\pi e}\geq \frac{1}{d^2} \frac{e^{2H_0}}{2\pi e},
\ee
which for a Gaussian prior of width $\sigma$ takes a simpler form 
\be
\mn{V(\omega)}\geq \frac{\sigma^2}{d^2}.
\ee

\subsection{Optimality of equatorial states}
\label{app:equator}

Consider any qubit state
\be
\ket{\psi_{\varphi}}= \cos(\varphi)\ket{0} + \sin(\varphi)\ket{1}
\ee
and a POVM element $\proj{{\bf n}}= \frac{1}{2}\left(\sigma_0 + {\bm \sigma}\cdot {\bf n}\right)$, we take $n_y=0$ without loss of 
generality for the argument. The probability to observe the outcome ${\bf n}$ given the evolution 
$U_\vartheta= e^{-\frac{\ii}{2} \vartheta \sigma_z}$ is
\bea\nonumber
p_\varphi({\bf n}|\vartheta)&=& \frac{1}{2}\left(1+ n_z \cos(2\varphi)  + n_x \sin(2\varphi) \sin{\vartheta}  \right)\nonumber\\
&=&\sin(2\varphi)\, p_{+}({\bf n}|\vartheta) + \frac{1}{2}\left( 1 - \sin(2\varphi) + n_z \cos(2\varphi) \right),\nonumber\\
\eea
where the subscript $+$ stands for $\varphi=\pi/4$ (and thus $\ket{\psi_{\pi/4}}=\ket{+}$). Accordingly, for any von Neumann 
measurement the statistics ${\bf p}_{\varphi}(\vartheta)=\left(p_\varphi({\bf n}|\vartheta), p_\varphi(-{\bf n}|\vartheta)\right)$ are 
given by
\be\label{eq:bad state}
{\bf p}_{\varphi}(\vartheta)= \sin(2\varphi) {\bf p}_{+}(\vartheta) + (1-\sin(2\varphi)) {\bf q}(\varphi, n_z),
\ee
where ${\bf q}(\varphi, n_z)$ is $\vartheta$-independent noise. The measurement statistics in Eq.\eqref{eq:bad state} can 
alternatively be reproduced as follows: one tosses a biased coin and in the case of heads (occurring with probability 
$\sin(2\varphi)$) runs the estimation with the state $\ket{+}$ yielding the statistics ${\bf p}_+(\vartheta)$, whereas in the case of 
tails one generates a random outcome from ${\bf q}(\varphi, n_z)$, and subsequently forgets about the result of the coin toss. 
From this construction it is obvious that it is always optimal to set $\sin(2\varphi)=1$,  which shows the optimality of the states on 
the equator for the estimation of the rotation generated by $\sigma_z$. Note also that this argument is not restricted to von 
Neumann measurements nor to any particular cost-function. 

\subsection{Single qubit solution for general prior}
 \label{app:general prior}
Here we give the solution of Eq.~(4) in the main text for a general prior $p_0(\omega)$.  For ease of exposition we use the notation 
$r_t = \text{Re}\left(\tilde p_0(t)\right)$ and  $i_t = \text{Im}\left(\tilde p_0(t)\right)$.  The optimal measurement operator is given by 
\begin{widetext}
\be
S_t=\left(
\begin{array}{cc}
 \frac{i_t r_t'-i_t' r_t}{2 \left(i_t^2+r_t^2-1\right)} & \frac{\left(-i_t^2+\ii r_t i_t+1\right) i_t'+\ii r_t' \left(r_t'+\ii i_t r_t-1\right)}{2 \left(i_t^2+r_t^2-1\right)} \\
 -\frac{\left(i_t^2+\ii r_t i_t-1\right) i_t'+r_t' \left(i_t r_t+\ii \left(r_t^2-1\right)\right)}{2 \left(i_t^2+r_t^2-1\right)} & \frac{i_t r_t'-i_t' r_t}{2 \left(i_t^2+r_t^2-1\right)} \\
\end{array}
\right),
\ee
which leads to 
\be
\tr\, \eta_t S_t = \frac{2\, i_t i_t' r_t r_t'+\left(i_t^2-1\right) (i_t')^2+\left(r_t^2-1\right) (r_t')^2}{2 \left(i_t^2+r_t^2-1\right)}.
\ee
\end{widetext}

\subsection{Approximative sequential strategy}
\label{app:sequential}

 Here we determine the optimal variance decrease of a sequential strategy under the assumption that the phase distribution 
 stays Gaussian after every measurement 
\be
 p_k(\omega) =\frac{1}{\sqrt{2\pi V_k}}e^{-\frac{\omega^2}{2 V_k}}.
 \ee
It is easy to link the variance at a step $k$, $V_k$,  to the variance at $V_{k+1}$. Given a measurement time $t_k$ for the prior 
$p_k(\omega)$ we get for $V_{k+1}$, using Eq.~(7),
 \be
 V_{k+1} = V_k (1 -V_k t_k^2 e^{-t_k^2 V_k}).
 \ee
Hence at each step of the protocol the variance is reduced by the factor $R_k \equiv(1 -A_k e^{-A_k})$ that only depends on the 
product $A_k=t_k^2 V_k$.   A natural choice is to require that this product is constant, and the same for all steps, $A_k=A$.  
Hence, $R_k = R = (1 -A e^{-A})$, which requires that the time duration grows exponentially with the number of steps 
$t_k^2 =A/V_k = \frac{A}{R^k} V_0$.  This choice then yields for the variance and total time at step $k$ the expressions
\bea
V_k&=& R^k V_0 \\
T_k&=& \sum_{\ell=0}^{k-1} t_\ell = \sqrt{\frac{A}{V_0}} \frac{R^{-k/2}-1}{R^{-1/2}-1}.
 \eea
 In particular one notes that asymptotically the precision scales quadratically with the total time 
 \be
 \left(1/V \right) \approx  \frac{\left(R^{-1/2}-1 \right)^2}{A} T^2,
 \ee
which is maximized by $ \left(1/V \right) \approx  0.08 \, T^2$ for $A\approx 0.63$.

\subsection{Degeneracy lifting with two spins-$1/2$}
\label{app:2spins}

\begin{figure}[!ht]
\centering
\includegraphics[width=0.98\columnwidth]{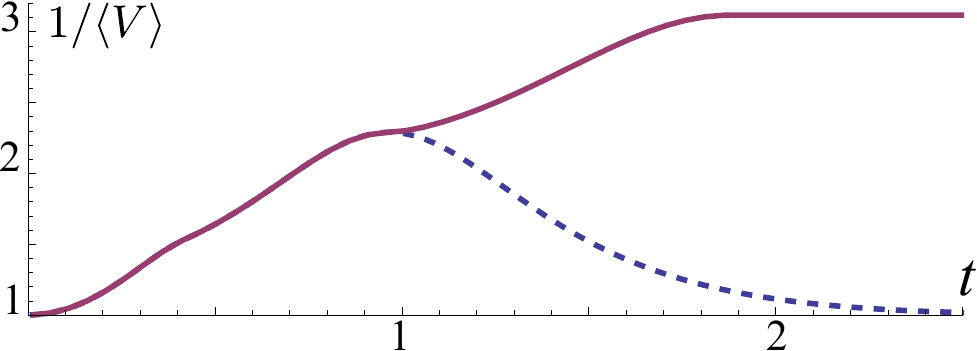}
\caption{The inverse of the final variance for the two qubits systems as a function of time. The solid line corresponds to the 
strategy where the degeneracy can be exploited with intermediate control pulses, 
whereas the dashed line corresponds to the free evolution without degeneracy lifting. The inset corresponds to the phases 
acquired by the various eigenvalues of the Hamiltonian leading to the best attainable precision.  Remark one can keep the optimal precision at the maximal level by simply 
freezing the phases of each level at the optimal point.}
\label{fig:2spins}
\includegraphics[width=0.98\columnwidth]{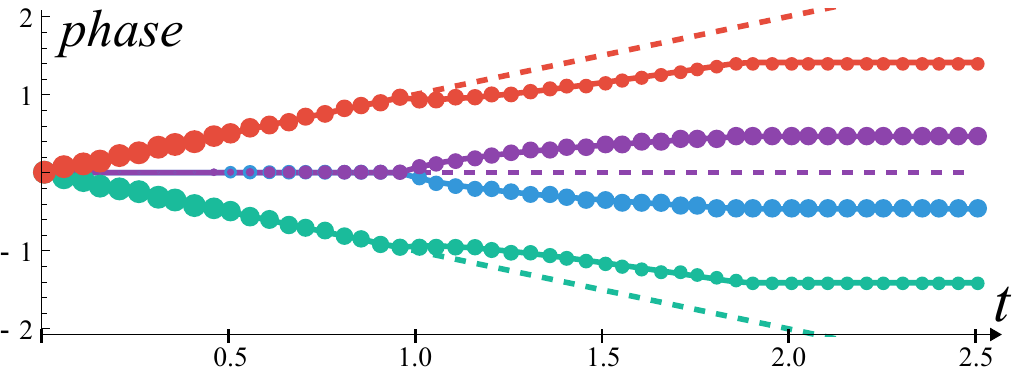}
\caption{The solid line depicts the optimal four-level spectrum that leads to the minimal MSE for each time. The area of each circle gives the weight of the corresponding level in the optimal state. The dashed line gives the free phase evolution.}
\label{fig:2spinssp}
\end{figure}

Given a Gaussian prior $p_0(\omega)=\frac{1}{\sqrt{2\pi}}e^{-\omega^2/2}$ consider the simplest non-trivial degenerate case of two spin-$1/2$ particles where the overall spectrum of the 
Hamiltonian is $\{+1,0,-1\}$ with a doubly degenerate $0$ eigenvalue. The corresponding eigenstates are $\{+1: \ket{\uparrow\uparrow},0: \ket{\downarrow\uparrow}$ and $\ket{\uparrow\downarrow} ,-1: \ket{\downarrow\downarrow}\}$. By performing intermediate Rabi flips between the states $\ket{\uparrow\uparrow}$ with $ \ket{\downarrow\uparrow}$, and  $ \ket{\downarrow\downarrow}$ with $ \ket{\uparrow\downarrow}$ it is possible to lift the degeneracy between the two $0$-levels (remark that this can be achieved with a simple $\pi$-pulse on the first qubit), but paying the price of slowing down the phase evolution of the outer levels $+1$ and $-1$. This becomes advantageous after a certain time as one can see from Fig.~\ref{fig:2spins} and  Fig.~\ref{fig:2spinssp}. In Fig.~\ref{fig:2spins} we compare the attainable precision for the case of a degenerate and non-degenerate spectrum. In Fig.~\ref{fig:2spinssp} we plot the optimal spectrum for each running time as compared to the free phase evolution. When time increases further the availability of four levels is not sufficient to avoid a phase wrap, the minimal MSE can however be preserved by freezing the phase of each level at the optimal value. This can be done by repeatedly applying Rabi flips between states $\ket{\uparrow\uparrow}$ with $\ket{\downarrow\downarrow}$, and $\ket{\downarrow\uparrow}$ with $\ket{\uparrow\downarrow}$. The optimal final state is then given by
\bea
\ket{\Psi}=  e^{ \ii\, 1.41 \,\omega} 0.42 \ket{\uparrow \uparrow} + e^{ \ii \,0.46 \,\omega } 0.57 \ket{\downarrow \uparrow}\nonumber\\+e^{- \ii \,0.46\, \omega } 0.57 \ket{\uparrow \downarrow}+e^{- \ii \,1.41 \,\omega} 0.42 \ket{\downarrow \downarrow},
\eea
and the optimal measurement reads 
\bea
S= \omega_1 \proj{\varphi_1^+}  - \omega_1 \proj{\varphi_1^-} + \nonumber\\ \omega_2 \proj{\varphi_2^+}  - \omega_2 \proj{\varphi_2^-} \\
\omega_1 \approx 1.38 \quad \omega_2 \approx 0.44 \nonumber\\
\ket{\varphi_1^\pm} \approx e^{\pm \ii\, 0.18} 0.38 \ket{\uparrow \uparrow} + e^{\pm \ii \,2.15} 0.59 \ket{\downarrow \uparrow}\nonumber\\+e^{\mp \ii \,2.15} 0.59 \ket{\uparrow \downarrow}+e^{\mp \ii \,0.18} 0.38 \ket{\downarrow \downarrow}\nonumber\\
\ket{\varphi_2^\pm} \approx e^{\mp \ii\, 0.99} 0.59 \ket{\uparrow \uparrow} + e^{\pm \ii \,0.18} 0.38 \ket{\downarrow \uparrow}\nonumber\\+e^{\mp \ii \,0.18} 0.38 \ket{\uparrow \downarrow}+e^{\pm \ii \,0.99} 0.59 \ket{\downarrow \downarrow}\nonumber
\eea

Observe that the optimal precision, state and measurement achieved by lifting the degeneracy of the Hamiltonian are very close to those for the non-degenerate, equally gapped Hamiltonian with four distinct eigenvalues, but this optimal precision is attained at a later time.  This is due to the fact that the spectral radius of the 
modified Hamiltonian is smaller than that of the $4$-level equally gapped one.


\subsection{Ion trap implementation}
 \label{app:ions}
 
We now describe a realization of our scheme for trapped ions. We make use of four internal electronic levels, 
$|g\rangle, |e\rangle$  which form the qubit, and two additional levels $|g'\rangle, |e'\rangle$, where the energy difference 
$E_{e'}-E_{g'} \not= E_{e}-E_{g}$. In addition, we consider the motion degrees of freedom that we label by 
$|k\rangle$, $1 \leq k \leq n$, where $k=1$ describes the motional ground state. The usage of such levels and their manipulation 
belongs to the standard repertoire of ion trap quantum information processing (see~~\cite{Alonso:13, Alonso:16, Ortiz:16}).

\begin{figure}[!ht]
\centering
\includegraphics[width=0.98\columnwidth]{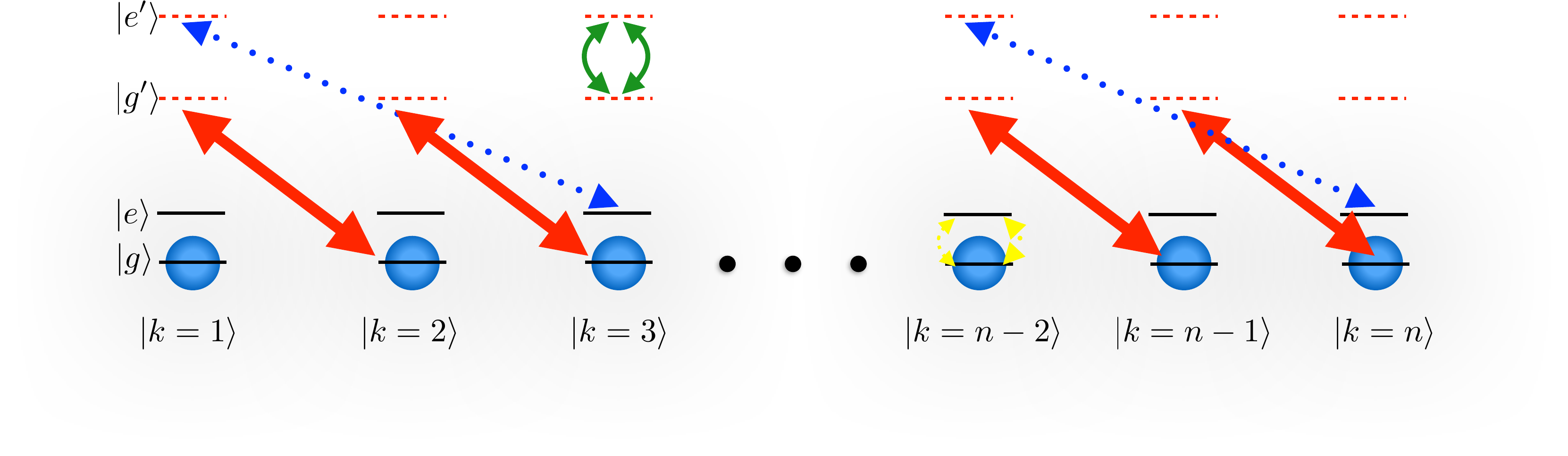}
\caption{The four main pulse sequences required to implement our degeneracy lifting protocol in an ion-trap setup.  The solid (red) arrows indicate the $G_1$ hiding pulses of Eq.~\eqref{pulses}, whereas the dotted (blue) arrows indicate the $E_2$ hiding pulses of Eq.~\eqref{pulses}.  The solid (green) and dotted (yellow) pulses indicated at $\ket{k=3}$ and $\ket{k=n-2}$ indicate the $X'$ and $\sigma_x$ pulses of Eq.~\eqref{puls} respectively. }
\label{ions}
\end{figure}

The protocol we describe here is a variant of the general protocol outlined in the main text, as the $n$ level system we will 
construct consists of only motional degrees of freedom.  The electronic degrees of freedom, while useful for the preparation of 
the state, will in the end act factor out.  We define the red de-tuned hiding pulses $G_l$ and $E_l$ that couple levels 
$|g'\rangle|j\rangle \leftrightarrow |g\rangle|j+l\rangle$ and $|e'\rangle|j\rangle \leftrightarrow |e\rangle|j+l\rangle$. That is, 
\bea\nonumber
G_l = \sum_j  |g'\rangle|j\rangle \langle g|\langle j+l| + h.c.\\
E_l = \sum_j  |e'\rangle|j\rangle \langle e|\langle j+l| + h.c. 
\label{pulses}
\eea
(see Fig.~\ref{ions}).  In addition, we consider the operators 
\bea\nonumber
X'=|g'\rangle\langle e'| + |e'\rangle \langle g'| \\
\sigma_x=|g\rangle\langle e|+ |e\rangle\langle g|, 
\label{puls}
\eea
which act as identity on the remaining levels. The evolution is governed by 
$H=B  \sigma_z =B (|g\rangle\langle g| - |e\rangle\langle e|)$. We remark that whilst an evolution of the states 
$|g'\rangle,|e'\rangle$ also takes place, this evolution can be safely ignored as we only populate these states at an intermediate 
stage for a short time. 

Our protocol consists of the following steps
\begin{itemize}
\item Prepare an arbitrary initial state of the form  $|\psi\rangle = |g\rangle \otimes \sum_k c_k |k\rangle$
\item At times $t_k=\frac{t(\lambda_k+1}{2}$ we transform $\ket{0, k}$ to $\ket{1,k}$ by applying to controlled operation
\be
U_k=\sigma_x\otimes|k\rangle\langle k| + \eins\otimes(\eins-|k\rangle\langle k|).
\label{flip}
\ee
This results in the final output state $|\psi\rangle = |e\rangle \otimes \sum_k c_k \, e^{-\ii B\lambda_k t}|k\rangle$.
\item Perform an arbitrary projective measurement on the motional degrees of freedom.
\end{itemize}
Notice the difference of this implementation to the general protocol described in the main text.  The electronic degrees of 
freedom are effectively coupled out, and the $n$-level system is comprised entirely of motional degrees of freedom.

We now show how to construct the controlled-flip operations $U_{k}$ of Eq.~\eqref{flip} by making use of a sequence of 
operations $G_lE_l X' G_lE_l$. Notice that the states $|g/e\rangle|k\rangle$ for $k<l$ are not affected by this operation, while for 
$k \geq l+1$ we obtain an operation $\sigma_x = |g\rangle\langle e|+ |e\rangle\langle g|$. It follows that $G_kE_k X' G_kE_k$ 
together with $G_{k-1}E_{k-1} X' G_{k-1}E_{k-1}$ implements $U_k$ for $k<n$, while for $k=n$ the operation 
$G_nE_n X' G_nE_n$ suffices. Note that $G_0E_0 X' G_0E_0$ can be replaced by $\sigma_x$.

What remains to be shown is the possibility to prepare arbitrary initial states 
$|\psi\rangle = |g\rangle \otimes \sum_k c_k |k\rangle$, and the performance of measurements on the resulting states 
(that are of the same form). A general operation on states $|\psi\rangle = |g\rangle \otimes \sum_k c_k |k\rangle$ can be 
obtained as follows. Observe that an arbitrary unitary operation $V^{(k_1k_2)}$ acting on any pair of levels 
$|k_1\rangle, |k_2\rangle$ can be realized by the sequence ${\cal M}=E_{k_2} U_{k_2} X' E_{k_1} U_{k_1}$, that transfers the 
states $|g\rangle|k_1\rangle,|g\rangle|k_2\rangle$ to auxiliary levels $|g'\rangle|1\rangle,|e'\rangle|1\rangle$, followed by an 
arbitrary operation $V'$ on the auxiliary levels $\{|g'\rangle,|e'\rangle\}$ (which can be realized in standard way). The auxiliary 
levels are mapped back via ${\cal M}$ to motional states $|k_1\rangle,|k_2\rangle$. Using sequences of operations 
$V^{(k_1k_2)}$ between pairs of levels, one can realize any unitary operation on the $n$-level system. This allows one to 
prepare an arbitrary initial state $|\psi\rangle$ from $|g\rangle|1\rangle$.

Finally, a projective measurement of the form $\{P_k=|k\rangle\langle k|,P^\perp\}$ can be achieved with $U_k$, followed by the 
standard projective measurement that couples the excited state $|e\rangle$ to some meta-stable auxiliary level via a laser pulse 
and detects the emitted photons. By a sequence of such (commuting) two-outcome measurements, a projective measurement in 
the basis $\{|k\rangle\}$ can be achieved. Together with general unitary operations that can be implemented as shown above, 
one can realize an arbitrary projective measurement.

We remark that for small $n$, more efficient schemes can be constructed where more of the electronic levels can be used. For 
instance, we can consider only motion states $|1\rangle$ and $|2\rangle$. In this case, we can make use of all four levels 
$|g/e\rangle|1/2\rangle$. We use $E_1$ to hide the state $|e\rangle|2\rangle$, then perform the red sideband pulse 
$|e\rangle|1\rangle \leftrightarrow |g\rangle|2\rangle$ and again $E_1$. By performing this pulse sequence at an appropriate 
time, one can generate an effective symmetric spectrum $\{-1, -\lambda_1, \lambda_1, 1\}$.  Also in this case, one can perform 
any operation on the four level using auxiliary levels $|g'\rangle,|e'\rangle$. Using $\sigma_x$ and 
$|g\rangle|1\rangle \leftrightarrow |e\rangle|2\rangle$, $|e\rangle|1\rangle \leftrightarrow |g\rangle|2\rangle$, together with $E_1$ 
and $X'$, one can transfer any two levels to the auxiliary system and perform an arbitrary two-level operation there. This allows 
one to prepare arbitrary initial states and perform arbitrary measurements on the four-level system.

\bibliographystyle{apsrev4-1}
\bibliography{degeneracy}
\end{document}